\documentclass[preprint,amsmath,amssymb,superscriptaddress,aip,apl,floatfix,showpacs]{revtex4-1}

\usepackage{graphicx}
\usepackage{dcolumn}
\usepackage{bm}
\usepackage{amsmath}

\begin{document}
\title{Strain-induced effects on band-to-band tunnelling and trap-assisted tunnelling in Si examined by experiment and theory}
\author{F. Murphy-Armando}
\email{philip.murphy@tyndall.ie}
\affiliation{Tyndall National Institute, Lee Maltings, Dyke Parade, Cork, Ireland}
\author{Chang Liu}
\affiliation{School of Electronic Science and Engineering, Nanjing University, 22 Hankou Rd, Nanjing 210093, China}
\author{Yi Zhao}
\affiliation{Department of Information Science and Electronic Engineering, Zhejiang University,38 Zheda Rd, Hangzhou 310027, China}
\author{Ray Duffy}
\affiliation{Tyndall National Institute, Lee Maltings, Dyke Parade, Cork, Ireland}
\date{\today{}}

\begin{abstract}
Strain is commonly used in metal-oxide-semiconductor technologies to boost on-state performance. This booster has been in production for at least a decade. Despite this, a systematic study of the impact of strain on off-state leakage current has been lacking. In this work we use experimental data and {\it ab-initio} calculations to refine existing models to account for the impact of strain on band-to-band tunnelling and trap-assisted tunnelling in silicon. We observe that the strain may dramatically increase the leakage current, depending on the type of tunnelling involved. For band-to-band and trap-assisted tunnelling, low uniaxial strains of 0.1\% (or 180 MPa) can increase the leakage current by 60\% and 10\% compared to the unstrained case, respectively. Using our models, we predict that compressive strain on the order of 1\% (or 2 GPa) can increase the leakage current by 150 times. Conversely, tensile strain may diminish or at most double the leakage current in all observed cases. Though detrimental in conventional inversion-mode MOSFETs, these processes may be used to boost the performance of Tunnel Field Effect Transistors, where on-state current is defined by band-to-band tunnelling.

\end{abstract}

\maketitle
\section{Introduction}
Leakage currents in metal-oxide-semiconductor (MOS) devices are undesirable as they drain power supply resources in integrated circuits and systems. The International Technology Roadmap for Semiconductors\cite{itrs}  (ITRS) specifies leakage targets for current and future generation MOS technologies, but it is not fully understood how difficult many of these targets will be to achieve, which physical mechanisms are most responsible, and what should be done to alleviate the expected problems. Literature on diode leakage has been available for many decades, however there are a number of aspects of modern MOS device processing and design that necessitates an update on the study of reverse biased junction leakage. 
Strain is considered mandatory for modern and future CMOS technologies, for on-state performance enhancement.\cite{song,ungersboeck}  This is achieved by an improvement of carrier mobilities in the device channel. Tensile strain is preferred for n-channel Si devices for electron mobility improvement, while compressive strain is better for p-channel Si devices. Strain can be introduced globally across the wafer, via epitaxial engineering of the substrate. Alternatively there are a number of local strain technologies including capping layers, such as SiN, or heteroepitaxial integration in the source and drain regions, such as SiGe in p-channel Si devices. However all these strain enhancements have focused on the on-state drive current in the MOS device, while relatively little study has been given to the impact of strain on the off-state. Off-state leakage current can originate at the reverse biased drain junction. Physical mechanisms to note here are band-to-band tunnelling (BBT), Shockley-Read-Hall (SRH) and trap-assisted tunnelling (TAT). 

In this work we measure the leakage current as a function of strain in several Si diode samples, selected to isolate the contribution from the different tunnelling mechanisms. We also wish to understand the leakage current under strain using available theory. It is our intention to use as few fitting parameters as possible. To accomplish this, we use $ab$ $initio$ electronic structure theory calculations to extract the electron-phonon coupling responsible for BBT and the strain effects on the band structure (effective masses, band gaps, etc.) and carrier populations. With the measured leakage currents and the $ab$ $initio$ model of BBT, we extract a best estimate of the trap lifetimes and energies, and their response to strain. The latter two are the only fitting parameters in our model, besides those related to avalanche breakdown.

We find that at stresses of up to 180 GPa the leakage current can be increased or reduced by the order of 5-10\%, and sometimes more at low voltages. Beyond the linear regime we predict that compressive strain can have a dramatic deleterious effect on the leakage current, leading to a 4 - 150 times increase for 1\%  uniaxial strain ($\sim$ 2 GPa), due to large increases in the BBT current. Interestingly, BBT is the mechanism that makes Tunnel Field Effect Transistors (TFETs) possible.\cite{appenzeller} 
While an increase in BBT current is harmful to conventional MOSFETs, it can enhance the performance of TFETs, and strain is already being explored to this effect.\cite{nayfeh}

\section{Experimental methods}
In a previous work,\cite{duffy1,duffy2} Si diodes were fabricated carefully with different doping concentrations in the substrate in order to isolate different leakage mechanisms, including BBT, SRH and TAT. This enabled a closer evaluation of these reverse leakage mechanisms.
In summary the process flow consisted of standard processing to define the active area and poly-buffered LOCOS isolation. A 5 nm screen oxide was deposited before the implants and anneals. For the n$^+$/p diodes high-concentration arsenic was implanted shallow and then given a high thermal budget anneal of 1100 $^\circ$C for 5 min to drive-in the dopant. Thereafter the low-concentration doped region was formed by a boron implant. The rest of the flow consisted of an 1100 $^\circ$C 0 s spike anneal to activate the dopants, a clean step to remove the screen oxide, and the deposition of a Ti/TiN contact layer. Finally a 650 nm layer of AlCu metal was deposited and patterned.
	Current-voltage characteristics of the fabricated diodes were measured using a HP4155 parameter analyzer. A thermochuck was used to investigate temperature dependency. For each structure and split several die were measured. Most of the measurements were done on square diodes, consisting of a rectangular active area with a high $\frac{area}{perimeter}$ ratio. For completeness perimeter current was extracted by comparing currents from meander structures with the same area but different isolation perimeters, and this was subtracted from the total square diode current. In all square diodes the perimeter component of the leakage current was several orders of magnitude lower than the area contribution. For details on the temperature dependence of the leakage current and the doping profile of the diodes we refer the reader to Refs. \onlinecite{duffy1} and \onlinecite{duffy2}.
	The mechanical stresses were applied using a four-point wafer bending system. Using this system, we studied the effects of both uniaxial tensile and compressive strain. Details of the mechanical stress equipment can be found in Refs. \onlinecite{yi1} and \onlinecite{yi2}.

\section{Refined model accounting for strain}

To account for the effects of strain on the leakage current, we modify the established models by A. Schenk\cite{schenk}  for BBT and G.A.M. Hurkx for SRH and TAT.\cite{hurkx3} The band structure of Si under strain is calculated using the 30-band k.p model of Rideau et al.\cite{kdp} The electron-phonon coupling necessary for the BBT is calculated using Density Functional Perturbation Theory, available in the code Abinit,\cite{abinit1,abinit2} as in Refs. \onlinecite{prb2,prberratum,jap2}. The total current density may be modelled by\cite{hurkx3}  

\begin{equation}\label{eq:jd}
j_d=\frac{\left(j_{bbt}\phi(0)+e \int_{-x_p}^{x_n} R_{trap}(x)\phi(x)\right) +j_{i} \left(1+ \phi(x_n)\right)/2}{1-\int_{-x_p}^{x_n}\alpha_n(x)\phi(x)dx}
\end{equation}
 				
where $\phi(x)$ is defined in Ref. \onlinecite{hurkx3} as
\begin{equation}
\phi(x)=e^{-\frac{1}{2}\int_{-x_p}^{x_n}\alpha_n(x')dx'},
\end{equation}
with the ionisation coefficient
\begin{equation}\label{eq:alpha}
\alpha_n(x)=\alpha_{n\infty}e^{-\frac{b_n}{\left|F(x)\right|}},
\end{equation}
where $F(x)$ is the electric field and $\alpha_{n\infty}$ and $b_n$ are fitting parameters related to avalanche breakdown.

The ideal diode current density $j_i$ is given by
\begin{equation}
j_i=j_s\left(e^{eV/kT}-1\right),
\end{equation}
where $j_s=e n_i^2\left(1/N_a\sqrt{D_n/\tau_n}+1/N_a\sqrt{D_p/\tau_p}\right)$ is the saturation current density, assuming a step doping profile, and $e$ is the electron charge, $n_i$ the intrinsic carrier concentration, $N_a$ and $N_d$ are the acceptor and donor densities, $D_n$ and $D_p$ the diffusion charge coefficients, and $\tau_n$ and $\tau_p$ are the  electron and hole lifetimes. At the carrier concentrations considered in this work, the contribution of $j_i$ to the total current is negligible.

The trap assisted tunnelling rate $R_{trap}$ is given by Eq. 6 of Ref. \onlinecite{hurkx2},
\begin{equation}\label{eqtat}
R_{trap}(x)=\frac{n(x)p(x)-n_i^2}{\tau \left[\frac{n(x)+n_i e^{\left(\overline{E}(x)\right)/kT}}{1+\Gamma_p(x)}+\frac{p(x)+n_i e^{\left(-\overline{E}(x)\right)/kT}}{1+\Gamma_n(x)}\right]},
\end{equation}
where $\overline{E}(x)=E_T-E_i(x)$, $E_T$ and $E_i$ are the average trap and intrinsic Fermi energies, respectively. $R_{trap}$ becomes the SRH tunneling rate when the field effect function\cite{hurkx2} $\Gamma \ll 1$ at low electric field.

In contrast to Ref. \onlinecite{hurkx2}, we consider the six conduction and three valence band valleys independently. Therefore we redefine the field effect functions
\begin{eqnarray}
\Gamma_n&=&\sum_c r_c \Gamma_{n,c}',\\
\Gamma_p&=&\sum_v r_v \Gamma_{p,v}',
\end{eqnarray}
where the sums are over the conduction bands at the six valleys at the $\Delta$ points in the Brillouin zone, 83\% along the $\Gamma-X$ line (index $c=\Delta_{\pm x},$ $\Delta_{\pm y},$ $\Delta_{\pm z}$) and three valence bands: heavy holes (HH), light holes (LH) and split-off band (SO) (index $v=HH$, $LH$, $SO$), $r_i$ are the fractional occupation of each band, and $\Gamma_{n,c}'$ and $\Gamma_{p,v}'$ are as defined in Eq. 7 of Ref. \onlinecite{hurkx2}, but for each band:
\begin{equation}
\Gamma_{n,c}'=\frac{\Delta E_{n,c}}{kT}\int_0^1 exp\left(\frac{\Delta E_{n,c}}{kT} u-K_{n,c}u^{3/2}\right)du
\end{equation}
\begin{equation}
K_{n,c}=\frac{4}{3}\frac{\sqrt{2 m_{n,c}^\parallel\Delta E^3_{n,c}}}{e\hbar \left|F\right|}
\end{equation}
and analogously for $\Gamma_{p,v}'$. The effective mass $m_{n,c}^\parallel$ corresponds to that in the transport direction for an electron in conduction band $c$, the $\Delta E_{n,c}$ and $\Delta E_{p,v}$ are given by Eqs. (9a) and (9b) of Ref. \onlinecite{hurkx2},
\begin{equation}
\Delta E_{n,c}=\begin{cases}
E_c(x)-E_{cn}, & E_T(x)\leq E_{cn}, \\
E_c(x)-E_T(x), & E_T(x)> E_{cn}
\end{cases}
\end{equation}
and
\begin{equation}
\Delta E_{p,v}=\begin{cases}
E_{vp}-E_v(x), & E_T(x)> E_{vp}, \\
E_T(x)-E_v(x), & E_T(x)\leq  E_{vp}.
\end{cases}
\end{equation}
$E_{c(v)}(x)=E_{c0(v0)}-\psi(x)$ is the local conduction band minimum (valence band maximum), and is given by the band structure at zero field and the local electrostatic potential, $\psi(x)$ defined below. $E_{cn}=E_c(x_n)$ and $E_{vp}=E_v(-x_p)$ are the conduction and valence band edges at the $n$ and $p$ sides, respectively. $x_{n,p}$ are the positions of the $n$ and $p$ sides, defined below.
\begin{figure*}
\includegraphics[width=\textwidth] {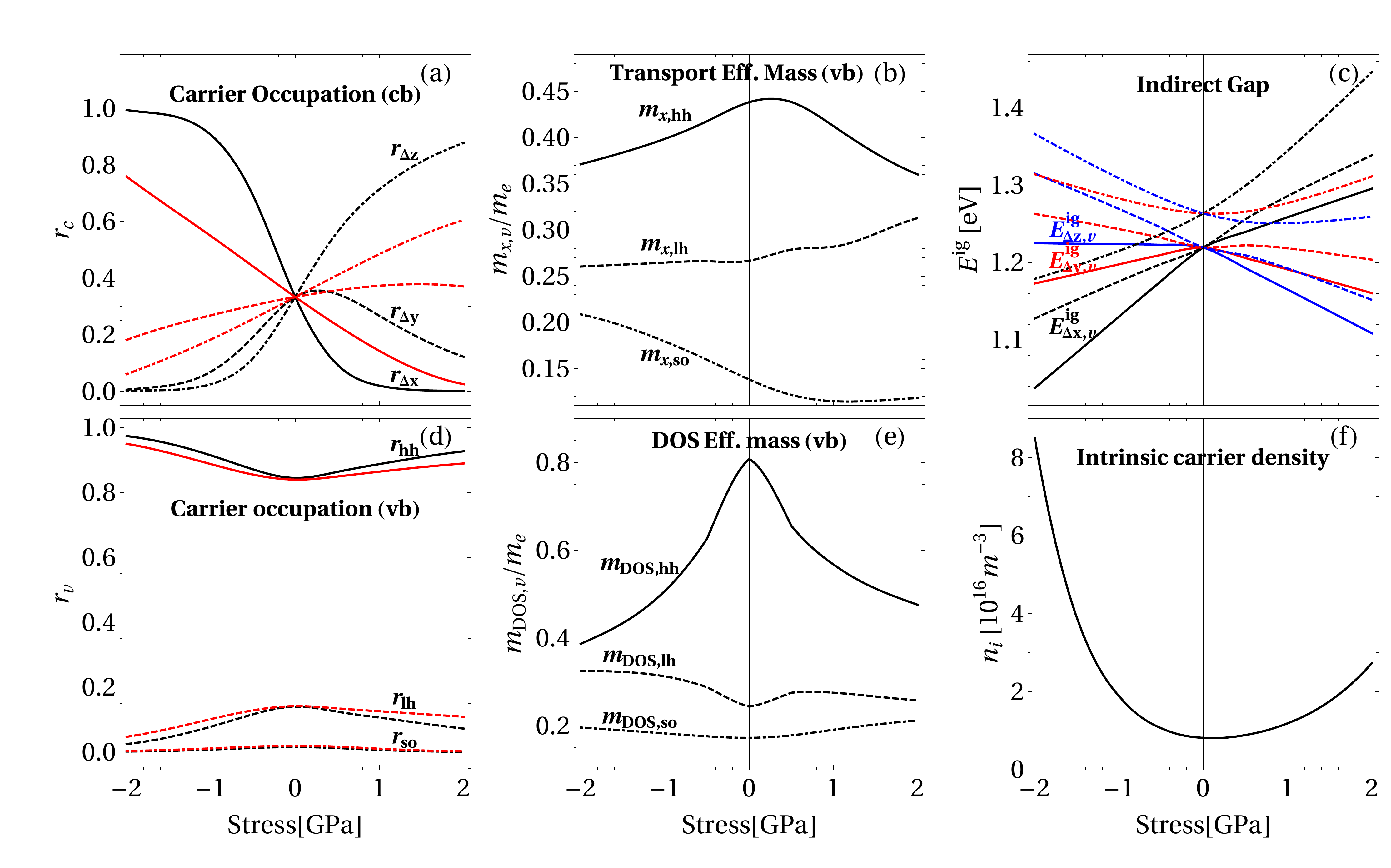}
\caption{\label{rcrvmxmdosegni2gpa} Theoretical (a) relative occupation of the conduction and (d) valence band valleys vs. strain for doping densities 1.9x10$^{18}$ cm$^{-3}$ (black) and 2.15x10$^{20}$ cm$^{-3}$ (red). The labels $\Delta$x, y and z refer to the $\Delta$ valleys along the three cartesian axes. The labels hh, lh and so refer to the heavy hole, light hole and split off valence bands of Si, respectively. (b) Conductivity mass vs. stress of the topmost valence bands in the $x$ direction. (c) Intrinsic energy gap vs. stress between the conduction band valleys (x, y and z valleys at $\Delta$ represented by black, red and blue, respectively) and the topmost valence bands (top: solid, second top: dashed, SO band: dot-dashed). (e) Density of states mass vs. strain of the three topmost valence bands. (f) Intrinsic carrier concentration vs. stress.}
\end{figure*}

 The coefficients $r_v$ and $r_c$ are new to the model, and give the proportional occupation of each of the valence ($v$) and conduction ($c$) bands, and depend on the amount of strain  (see Figs. \ref{rcrvmxmdosegni2gpa}(a) and  \ref{rcrvmxmdosegni2gpa}(d) for the strain dependence of $r_c$ and $r_v$). All parameters have been calculated as strain-dependent.

\begin{figure}
\includegraphics[width=3in] {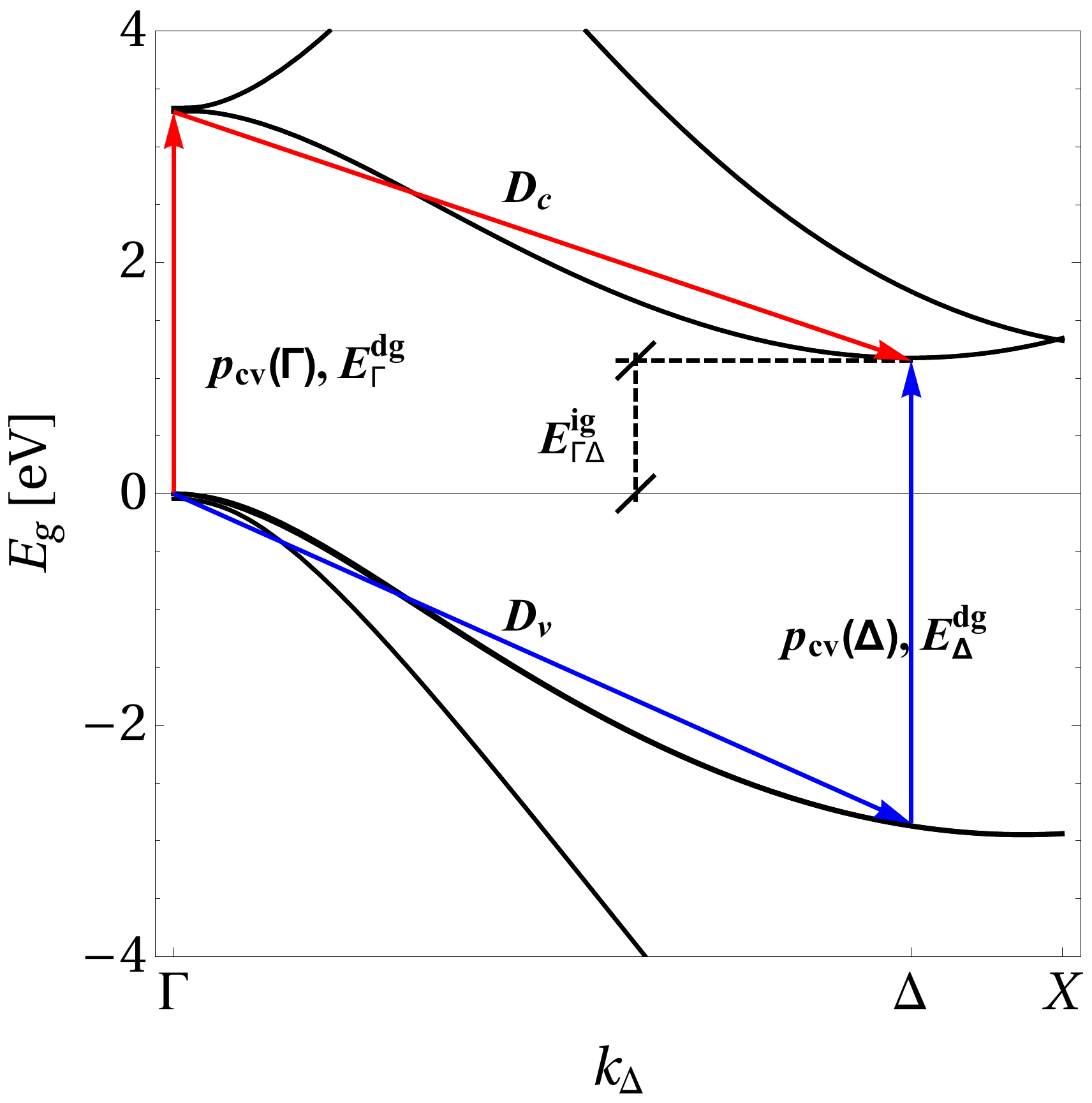}
\caption{\label{elphch} Theoretical electron-phonon channels considered in BBT, shown on the electronic band structure of Si.}
\end{figure}

Based on the model by Schenk,\cite{schenk} the BBT current is given by

\begin{eqnarray}\label{eqbbt}
j_{bbt}&=&\frac{12\pi e^2}{\hbar^5}\sum_{i,c,v} r_c r_v \times \nonumber \\
&\times& \left(\left| V_{c,i}^e\right|^2 \left|\frac{p_{c,v}(\Gamma)}{E_{c,v}^{dg}(\Gamma)}\right|^2
+\left| V_{v,i}^p\right|^2 \left|\frac{p_{c,v}(\Delta)}{E_{c,v}^{dg}(\Delta)}\right|^2\right) \times \nonumber \\
&\times& \frac{\sqrt{m_t^2 m_l}m_{DOS,v}^{3/2}}{m_e^2} F_m [\left(E_{c,v}^{ig} -\hbar\omega_i\right)N_i H(x_{i,c,v}^-)+\nonumber \\
&&+ \left(E_{c,v}^{ig} +\hbar\omega_i\right)\left(N_i+1\right) H(x_{i,c,v}^+) ] \times \nonumber \\
&\times &\left(f_v(\Gamma)-f_c(\Delta)\right),
\end{eqnarray}

\begin{equation}
H(x)=\frac{Ai(x)}{x^2}+\frac{Ai'(x)}{x}+Ai_1(x),
\end{equation}
\begin{equation}
x_{i,c,v}^\pm=2^{2/3}\left(E_{c,v}^{ig} \pm \hbar \omega_i\right)\left(\frac{2 \mu_{c,v}}{e^2\hbar^2F_m^2}\right)^{1/3},
\end{equation}
where $Ai(x)$, $Ai’(x)$ and $Ai_1(x)$ are Airy functions, $\pm$ corresponds to emission ($+$) and absorption ($-$) of a phonon, $V_{c,i}^e$  and $V_{v,i}^p$  are the electron- and hole-phonon scattering potentials between states at $\Gamma$ and $\Delta$, at the conduction and valence band, respectively, $p_{c,v}(\Gamma$ or $\Delta)$ are the optical valence-conduction coupling constants at $\Gamma$ and $\Delta$. The terms involving  $V_{c,i}^{e/p}$ and $p_{c,v}(\Gamma$ or $\Delta)$ involve the two processes shown in Fig. \ref{elphch} between the valence and conduction bands via the dipole and electron-phonon matrix elements. It is analogous to the the process of indirect photon absorption, where the electric field coupling is given by that of the photon, rather than the field across the junction.

The masses $m_t$, $m_l$ and $m_{DOS,v}$  correspond to the transverse and longitudinal masses at the conduction band at $\Delta$, and the density-of-states (DOS) mass of the valence band at $\Gamma$. 
$E_{c,v}^{ig(dg)} $ is the indirect (direct) band-gap between the conduction band $c$ and valence band $v$ (see Fig. \ref{elphch}), $\omega_i$ is the frequency of phonon mode $i$ involved in the transition, $N_i$ is the number of phonons in mode $i$,  given by the Bose-Einstein distribution, and $f_n(m)$ is the electronic distribution of band $n$ at the $m$ k-point ($\Gamma$ or $\Delta$). The reduced mass $\mu_{c,v}$ is given by the conductivity masses in the direction of the current
\begin{equation}
\frac{1}{\mu_{c,v}}=\frac{1}{m_v^\parallel}+\frac{1}{m_c^\parallel},
\end{equation}
where $m_c^\parallel$ depends on the conduction band valley, and is given by $m_{cx}^\parallel=m_{cy}^\parallel=0.19 m_e$ and $m_{cz}^\parallel=0.95 m_e$, and $m_v^\parallel$ is given below in Eqs. \ref{eq:mdos} and \ref{eq:mv}.

For simplicity, we calculate the maximum electric field $F_m$, and the depletion width $W$ in terms of the applied voltage by assuming a step doping profile. We have also performed the calculations with the potential given by the doping profile measured in Refs. \onlinecite{duffy1} and \onlinecite{duffy2}, but found little difference. The step doping profile is given by the potential,
\begin{equation}
\psi(x)= 
    \begin{cases}
      \frac{eN_a}{2k_0}\left(x+x_p\right)^2, &  x<-x_p \\
      V_d-V-\frac{eN_d}{2k_0}\left(x-x_n\right)^2, & -x_p\leq x \leq x_n \\
      V_d-V, &x>x_n
    \end{cases}
\end{equation}
that results in an electric field,
\begin{equation}
F(x)=\begin{cases}
F_m \left(1-\frac{\left|x\right|}{x_p}\right),   & x<0 \\
F_m \left(1-\frac{\left|x\right|}{x_n}\right),  &   x\geq 0
\end{cases}
\end{equation}
where the coordinates for the pure $n$ and $p$ sides are given by
\begin{equation}
x_{n,p}=\sqrt{\frac{2N_{a,d} k_0 V_d \left(1-V/V_d\right)}{eN_{d,a} \left(N_d+N_a\right)}}.
\end{equation}
The maximum electric field is
\begin{equation}
F_m=2 \frac{V_d}{W}\left(1-\frac{V}{V_d}\right)^\frac{1}{2},
\end{equation}
with the junction width given by
\begin{equation}
W=\sqrt{\left(2 k_0/e\right)\left(\frac{1}{N_d}+\frac{1}{N_a}\right)\left(V_d-V\right)},
\end{equation}
and the junction voltage $V_d$ given by
\begin{equation}
V_d= E_{Fn}-E_{Fp},
\end{equation}
where $E_{Fn}$ and $E_{Fp}$ are the quasi Fermi energies in the $n$ and $p$ sides respectively.
The DOS and conductivity masses at the valence band are parametrized from the total band structure as an average in the occupation of holes:\cite{fisch}
\begin{equation}\label{eq:mdos}
m_{DOS,v}^{3/2}=\frac{\pi^2 \hbar^3}{\sqrt{2}\left(kT\right)^{3/2}}\frac{n_v}{\int \frac{\sqrt{x}}{1+e^{x-\left(E_v-E_F\right)/kT}}dx},
\end{equation}
and
\begin{equation}\label{eq:mv}
m_v^\parallel=\left(\int \frac{1}{m_v({\bf k})}\left(1-f_v({\bf k})\right){\bf dk}\right)^{-1}
\end{equation}
respectively, with $E_v$ the energy of valence band $v$,  and $E_F$ the Fermi level. See Figs. \ref{rcrvmxmdosegni2gpa}(b) and \ref{rcrvmxmdosegni2gpa}(e) for the stress dependence of the current and DOS masses, respectively.
\begin{figure}
\includegraphics[width=4in] {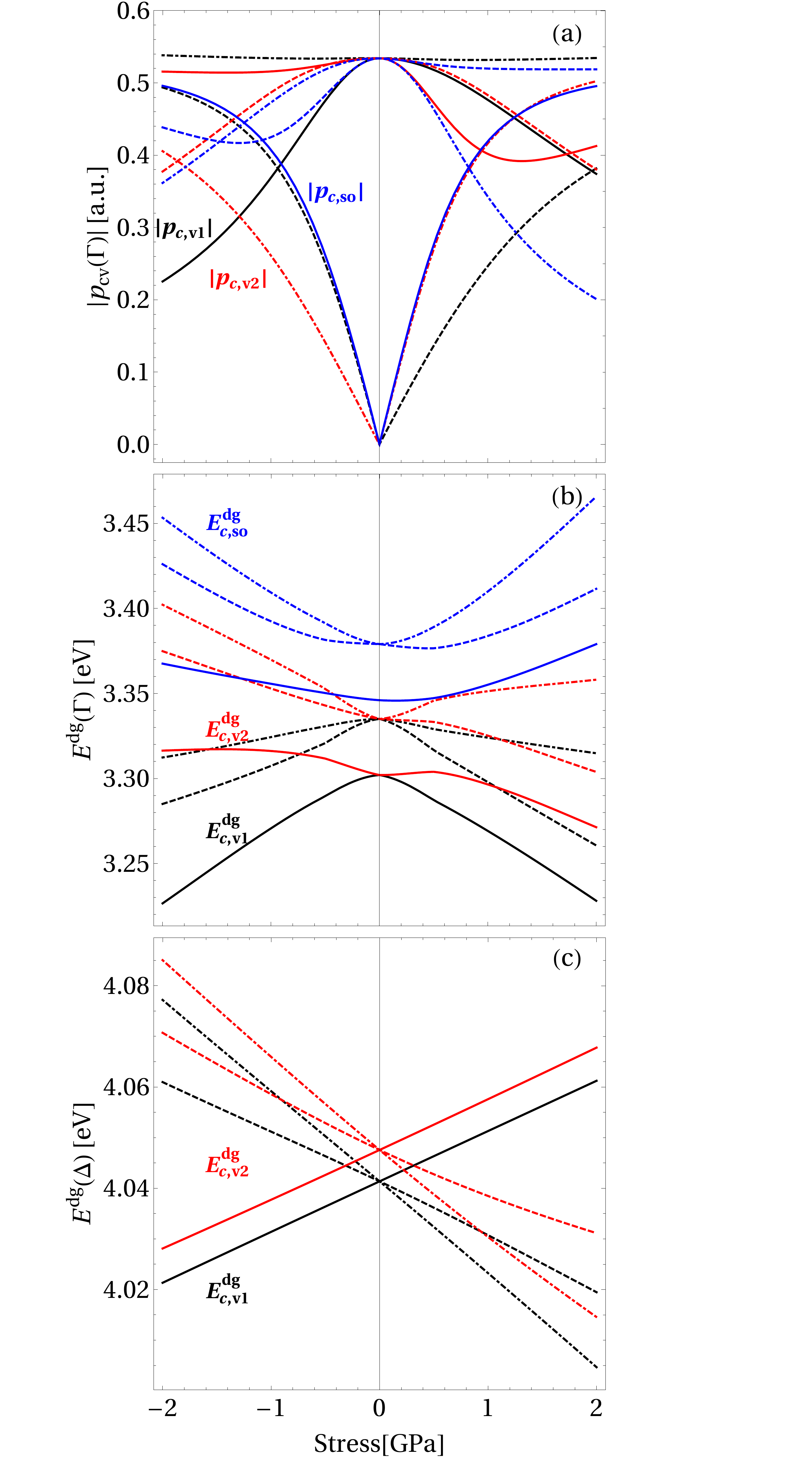}
\caption{\label{pcvedgedd} Theoretical (a) dipole matrix element, (b) direct band gap at $\Gamma$ and (c) direct band gap at $\Delta$ vs. stress between the first three conduction bands at $\Gamma$ (with increasing energy represented by solid, dashed and dot-dashed) and the three topmost valence bands (top: black, second top: red, SO split off band: blue). The direct band gap $E_{c,so}^{dg}(\Delta) \gg E_{c,(v1,v2)}^{dg}(\Delta)$, resulting in a negligible term in $j_{bbt}$, and is therefore not shown.}
\end{figure}

All parameters in the BBT current density are determined {\it ab initio}. The inter-valley electron-phonon coupling is given by the inter-valley deformation potentials as
\begin{equation}
\left|V^{e/p}_{c/v,i}\right|^2=D_{c/v}^2\frac{\hbar}{2 \rho \omega_i},
\end{equation} 
where $D_{c/v}$ is the inter-valley deformation potential for a phonon linking the $\Gamma$ and $\Delta$ at the conduction/valence band, $\rho$ is the atomic density and $\omega_i$ is the frequency of phonon with polarisation $i$. The calculated deformation potentials and frequencies are listed in Table \ref{param}. The dipole matrix elements between valence and conduction band are determined from the k.p model hamiltonian $H_{\bf k}$ as $p_{c,v}=\left\langle \phi_c \right|\bigtriangledown_{\bf k}H_{\bf k}\left|\phi_v\right\rangle$. The dipole matrix elements and direct band gaps vs. stress at the $\Gamma$ k-point are shown in Fig. \ref{pcvedgedd} (a) and (b), respectively. The dipole matrix elements at the $\Delta$ k-point are stress independent with values for $\left| p_{\Delta c,v1}\right|=\left| p_{\Delta c,v2}\right|=0.57$ and $\left| p_{\Delta c,so}\right|=0$, in atomic units. The direct band gaps vs. stress at the $\Delta$ k-point are shown in Fig. \ref{pcvedgedd} (c).

\section{Fitted quantities}
\begin{figure}
\includegraphics[width=3.4in] {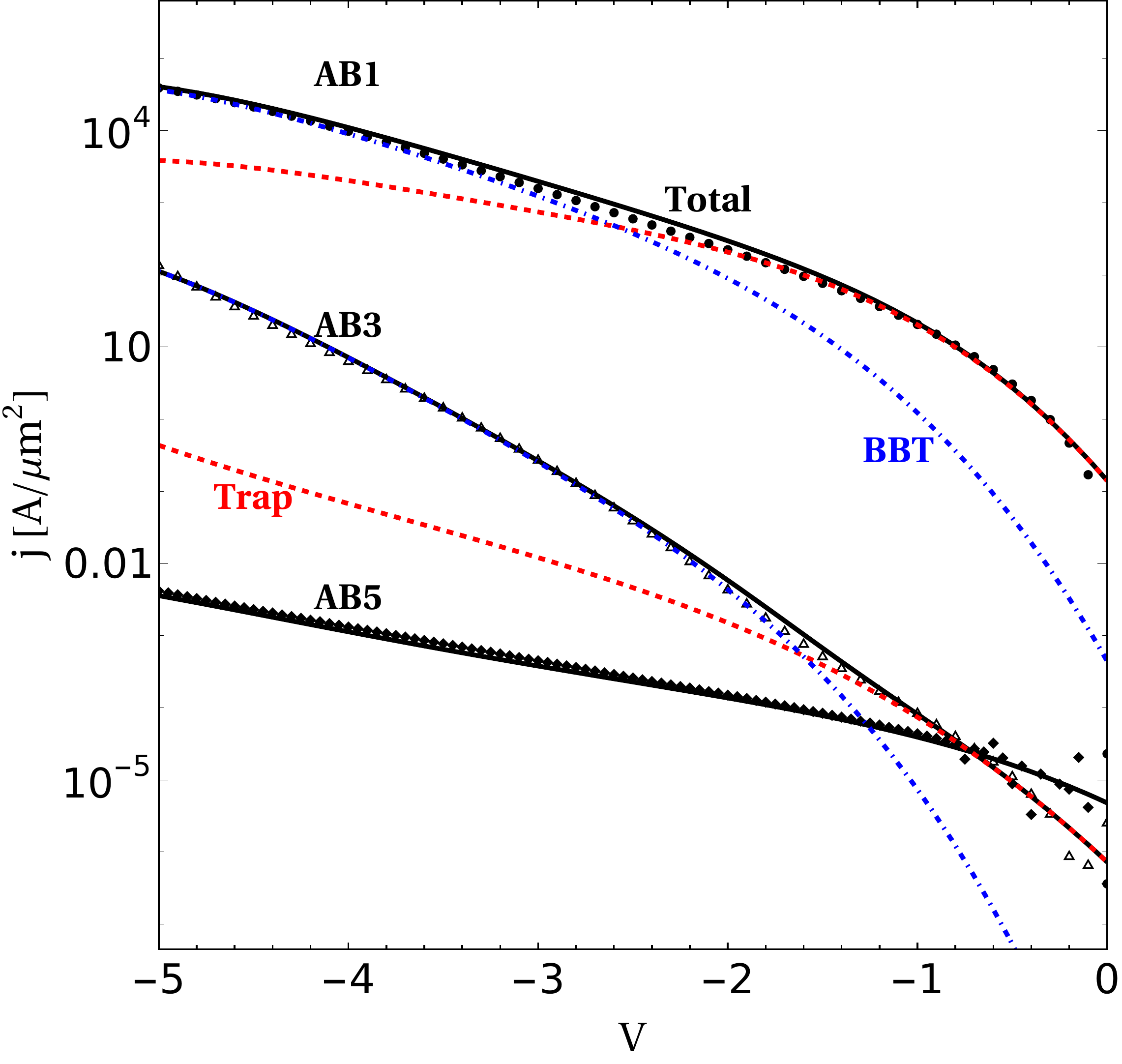}
\caption{\label{initialpnab} Experimental (points) and modelled (solid) current density vs voltage for the 3 samples considered. The modelled current is composed of band-to-band tunnelling (blue, dot-dashed), trap-assisted-tunnelling (red, dashed) and total current (black solid).}
\end{figure}

In Eq. \ref{eqbbt}, the only free parameter affecting the BBT current is the doping concentration. Therefore, we determine the effective densities of acceptor impurities $N_a$ for samples AB1 and AB3 at zero strain by the BBT dominated current at high $\left|V\right|$. In AB5 we determine $N_a$ together with the trap lifetime $\tau$ and trap energy $E_T$. The effective impurity densities $N_a$ for all samples are shown in Table \ref{fitp}. All agree within a factor of 2  with the maximum experimentally determined doping densities at the junction\citep{duffy1} of $6.1\times 10^{18}$, $1.9\times 10^{18}$ and $1\times 10^{17}$ for AB1, AB3 and AB5 respectively. Figure \ref{initialpnab} shows the measured and modelled current densities at zero strain for the samples considered.

Many of the parameters involved in TAT are very hard to calculate, as the number, energy and lifetimes of the traps are unknown. We have therefore fitted the lifetime $\tau$ and the trap energies $E_T$ to our experiments. From Eq. \ref{eqtat} we can see that $\tau$ is a scaling factor at all voltages. The effect of $E_T$ is more complicated, in that it changes the slope of the current vs. voltage. Therefore, $E_T$ has a large effect on the response to strain, affecting the sign of the slope of the current vs. strain. 
\begin{figure}
\includegraphics[width=3.9in] {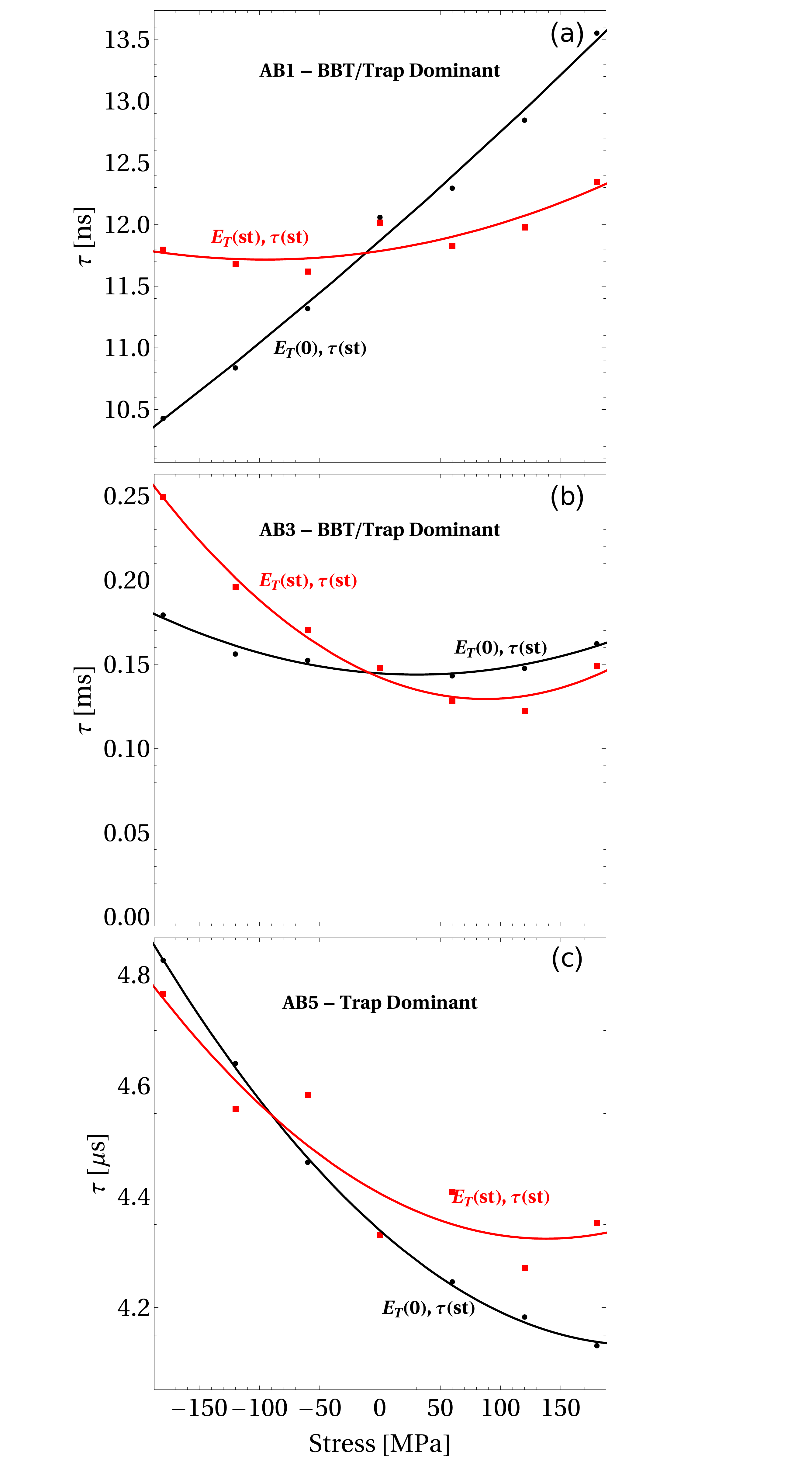}
\caption{\label{tauvstr}  Fitted trap lifetime $\tau$ vs strain for samples (a) AB1, (b) AB3 and (c) AB5. The fitted values depend on whether the trap energy $E_T$ is strain dependent (red) or not (black). The currents corresponding to these values are shown in Figs. \ref{jvstm1v} and \ref{jvstm3v}.}
\end{figure}

\begin{figure}
\includegraphics[width=3.9in] {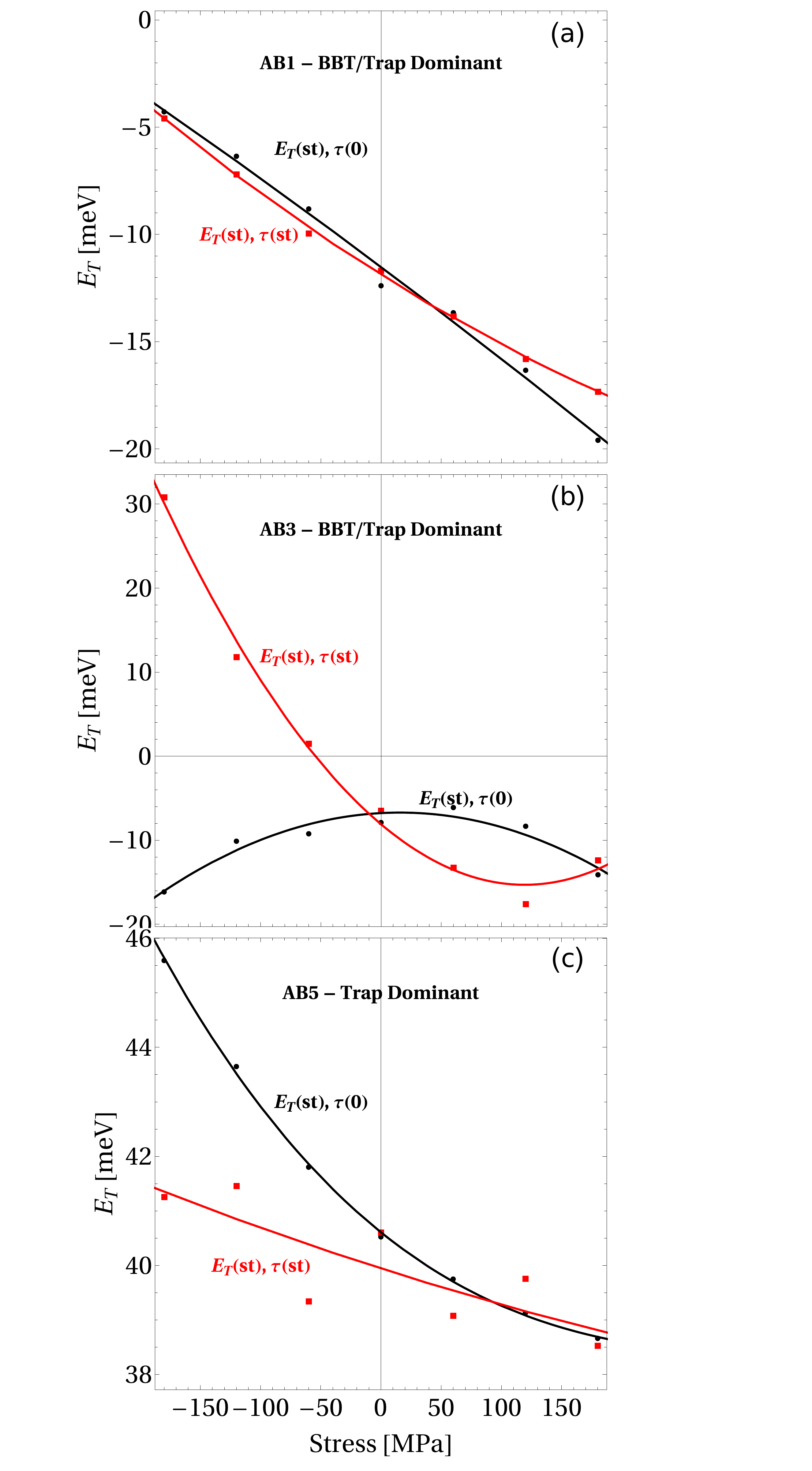}
\caption{\label{etvstr} Fitted trap energies $E_T$ vs strain for samples (a) AB1, (b) AB3 and (c) AB5. The fitted values depend on whether the trap lifetimes $\tau$ are strain dependent (red) or not (black). The currents corresponding to these values are shown in Figs. \ref{jvstm1v} and \ref{jvstm3v}.}
\end{figure}

We considered trap lifetime $\tau$ and trap energy $E_T$ to be strain dependent and voltage independent. The best fit to the measured current vs. stress is given by a cubic dependence on stress. The resulting fits for $\tau$ and $E_T$ vs. stress are shown in Figs. \ref{tauvstr} and \ref{etvstr}, respectively. 

The values for $\tau$ and $E_T$ are meant to reflect the effective values for an unknown variety of trap states. We believe that the cubic dependence at higher stresses is unlikely, and rather expect a saturation or at least a linear behaviour. We therefore consider three further limiting cases for the stress effect on the trap assisted current: both or either $\tau$ and $E_T$ constant with stress. The resulting values considering these cases for the fitted  $\tau$ and $E_T$ with stress are shown in Figs. \ref{tauvstr} and \ref{etvstr}.


\begin{table}
\caption{\label{fitp} Fitted parameters used in Eqs. \ref{eq:jd} and \ref{eq:alpha}.
}

\begin{ruledtabular}
\begin{tabular}{ l c c c }

&AB1&AB3&AB5\\
\hline
$N_a$ $\left[10^{17}cm^{-3}\right]$ & 35.63 & 14.82 & 2.5\\
$a_{n\infty}$ [m$^{-1}$]	& 1 & $1.824\times 10^{7}$ &	$6.9\times 10^{6}$ \\
$b_n$ [V/m]	&1 & $3.0\times 10^{5}$ &	1\\

\end{tabular}
\end{ruledtabular}
\end{table}

The electron-phonon process responsible for BBT is related to that causing indirect photon absorption and emission in Si. This has been explored from first principles in a previous work,\cite{noff} however the deformation potentials have not been published, and hence we present ours here in Table \ref{param}. 
Vandenberghe and Fischetti\cite{vandenberghe} have also studied BBT using ab initio calculated the electron-phonon coupling parameters. However, they considered a direct valence-to-conduction electron-phonon route, rather than one mediated by the electric field, as in the analysis by A. Schenk\cite{schenk} (see Figure \ref{elphch}). 
The direct valence-to-conduction electron-phonon route would only be valid for overlapping valence and conduction band energies to within the phonon energy. The coupling resulting from the latter channel would be independent of the electric field, except indirectly through the change in the valence and conduction band energies. We believe this channel would have a very small contribution at the electric fields considered in this work, and is therefore neglected.
We present the calculation of the current densities with parameters calculated entirely ab initio, and find good agreement to our measured currents in BBT dominant samples at zero strain (see Fig. \ref{initialpnab}).

\begin{table}[!h]
\caption{\label{param} Calculated electron-phonon deformation potentials and phonon energies for the phonons involved in BBT.}

\begin{ruledtabular}
\begin{tabular}{ l c c r }

Phonon&$\hbar \omega$ [meV]&$D_c$[eV/\AA]&$D_v$ [eV/\AA]\\
\hline
2TA	& 17 &	0.15 &	0.28 \\
LA	& 42 &	4.6 &	4.9\\
LO	& 57 &	0 &	7.2\\
2TO	& 57 &	5.4 &	2.7\\

\end{tabular}
\end{ruledtabular}
\end{table}

\section{DISCUSSION}
\begin{figure*}
\includegraphics[width=\textwidth] {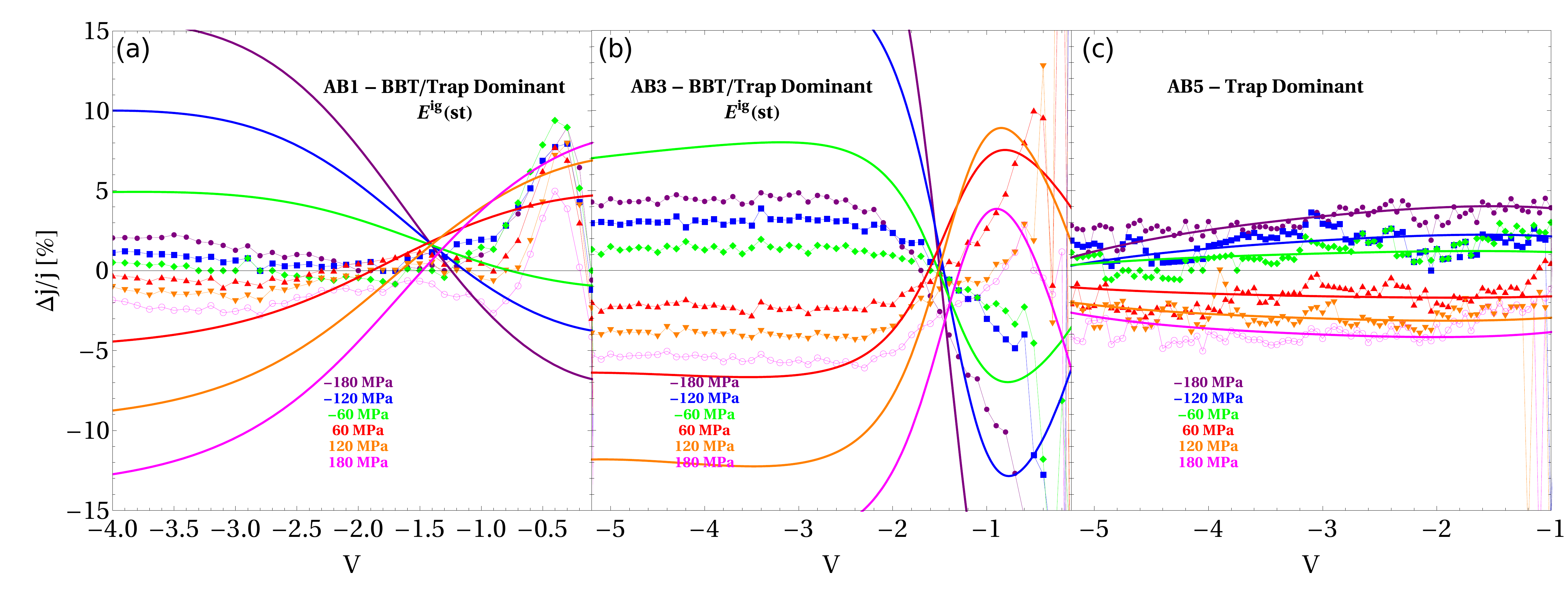}
\caption{\label{djAB135vst1-4vc} Experimental (joined points) and modelled (solid) current density change (in percent) vs reverse bias voltage for samples (a) AB1, (b) AB3 and (c) AB5 at 6 different stresses  (-180, -120, -60, 60, 120, 180) MPa represented as (purple, blue, green, red, orange and magenta).}
\end{figure*}

We measured and calculated the reverse bias current percent-change along the (001) direction as a function of uniaxial stress in the (100) direction in all the samples. Figure \ref{djAB135vst1-4vc} shows these for samples AB1, AB3 and AB5, respectively. For such small stresses, the change in current density effected by the strain is quite large, up several percent for strains of up to 0.13\% (see also results for BA2 in Ref. \onlinecite{this}). At voltages $-0.5<V<0$ the measured change of the current with stress can be of several 10s of percent (not shown).
From Eqns. \ref{eqtat} and \ref{eqbbt} we expect the behaviour of TAT and BBT currents to be very different under applied strain. From our model, samples AB1 and AB3 are dominated by TAT between 0 and -1.5V, and BBT thereafter (see Figure \ref{initialpnab}). The crossover from TAT to BBT at -1.5 V can also be clearly seen in the strain response shown in Fig. \ref{djAB135vst1-4vc}.

\begin{figure*}
\includegraphics[width=5in] {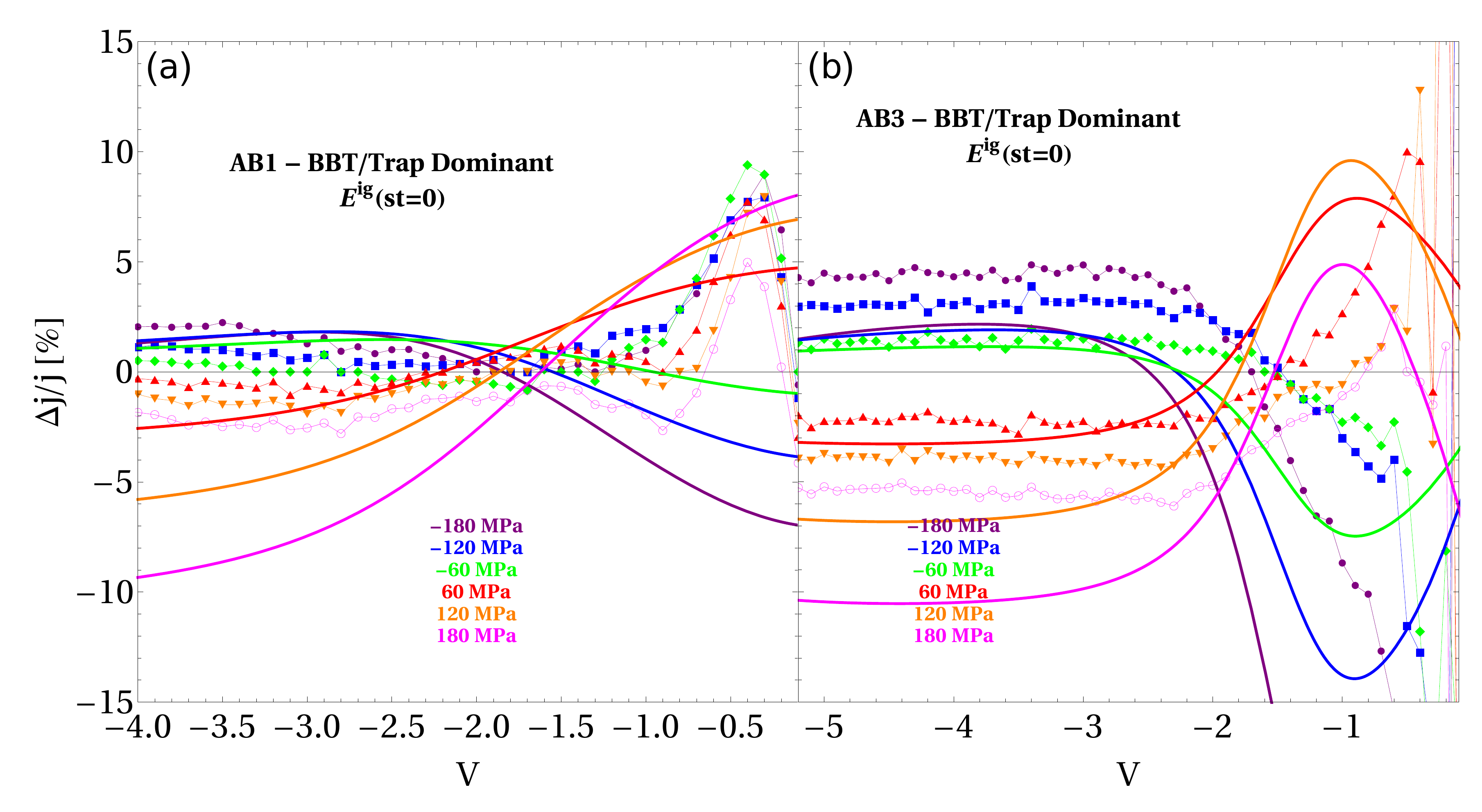}
\caption{\label{djAB135vst1-4vc-egi}Experimental (joined points) and modelled (solid) current density change (in percent) vs reverse bias voltage neglecting the strain dependence of $E^{ig}$ for samples (a) AB1, (b) AB3 and (c) AB5 at 6 different stresses  (-180, -120, -60, 60, 120, 180) MPa represented as (purple, blue, green, red, orange and magenta).}
\end{figure*}

Our model reproduces the current vs. voltage behaviour very well, but overestimates the response to strain. This may be due to several reasons, such as reduced transfer of stress into the sample, or the effect of very high doping on the strain response of the band gaps. In our calculations, we did not include any effects of the doping on the band gaps and effective masses. The BBT is exponentially sensitive to variations in the indirect band gap, and doping may be reducing its change with strain. This effect is clearly seen in Fig. \ref{djAB135vst1-4vc-egi}, which shows the effects of neglecting the strain dependence of the indirect band gap $E^{ig}$ on the current response to stress in eq. \ref{eqbbt}. A stress independent $E^{ig}$, while keeping all other stress dependencies, produces excellent agreement to experiment. This suggests that the overlap of the decaying electron and hole wavefunctions at each side of the junction is not changing with stress as fast as in the model.
We expect, however, that the sensitivity of $E^{ig}$ to strain to increase at higher strains, as the higher energy separation between valleys cannot be compensated by the doping charge. Measurements at higher strains and further study into the effects of doping on the band gap in the presence of strain are required to answer these questions.

\begin{figure}
\includegraphics[width=3.2in] {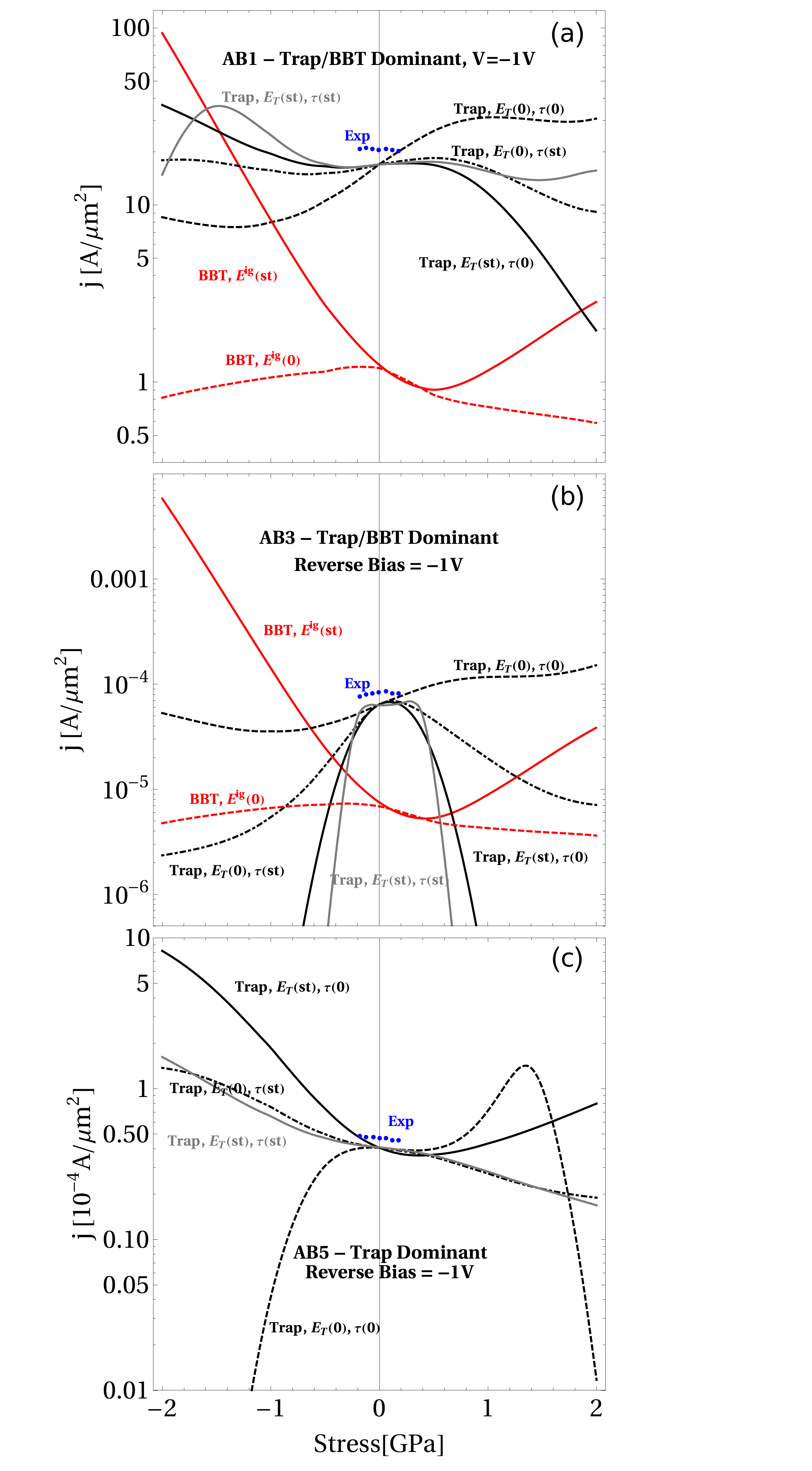}
\caption{\label{jvstm1v} Contributions from the currents of samples (a) AB1, (b) AB3 and (c) AB5 due to BBT (red) and Trap (black and grey) vs applied stress at a reverse bias voltage of $-1$ V. The current represented by the dashed BBT curve neglects the strain contribution to the indirect gap $E^{ig}$. The solid BBT curve includes the full strain dependence. The Trap assisted currents are calculated with full strain dependence of $E_T$ and $\tau$ (grey), only $E_T$ (solid), only $\tau$ (dot-dashed) and strain independent $E_T$ and $\tau$ (dashed) (see Figs. \ref{tauvstr} and \ref{etvstr} for the strain dependence of $\tau$ and $E_T$). The blue points correspond to the experimentally measured current density.}
\end{figure}

\begin{figure}
\includegraphics[width=3.2in] {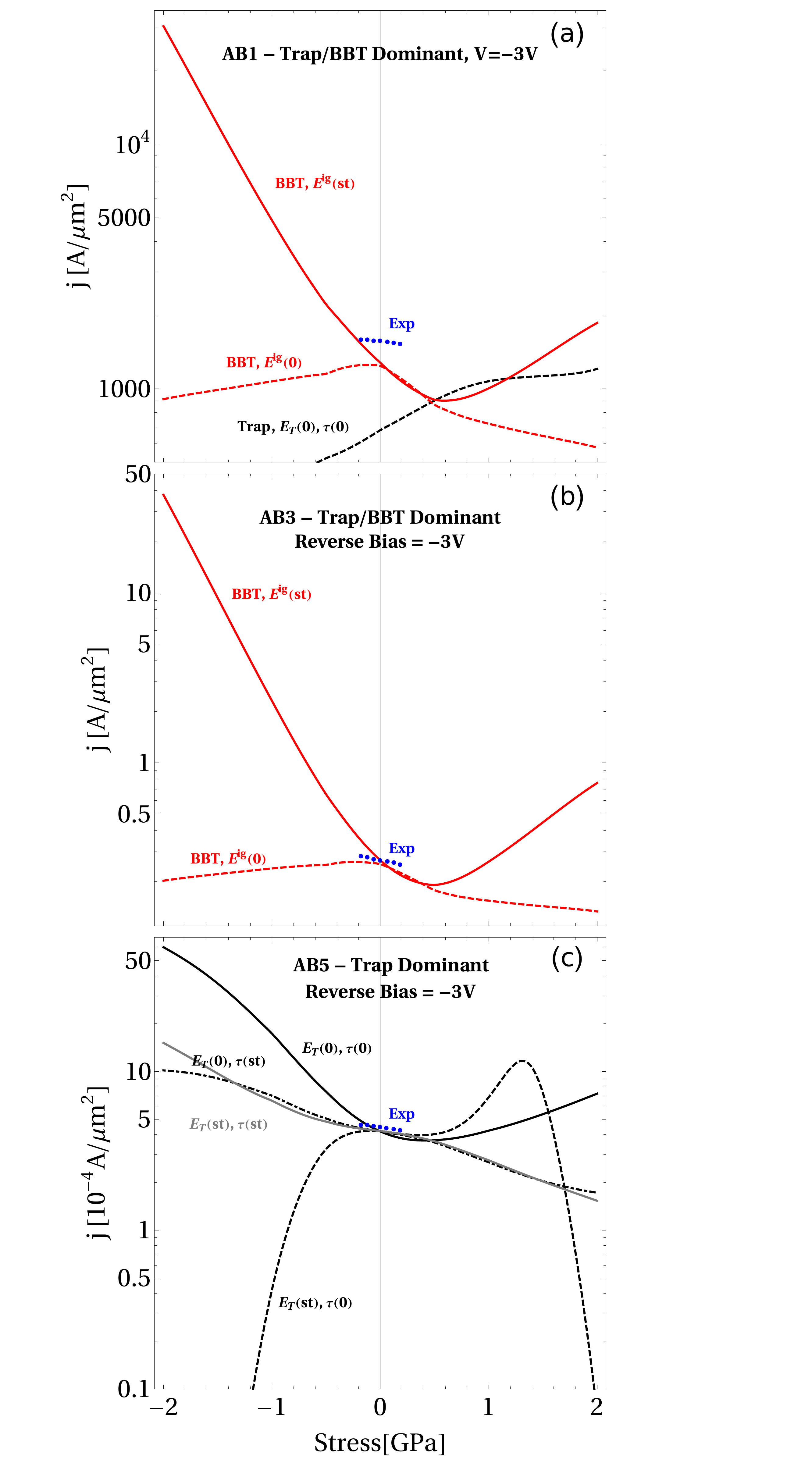}
\caption{\label{jvstm3v} Contributions from the currents of samples (a) AB1, (b) AB3 and (c) AB5 due to BBT (red) and Trap (black) vs applied stress at a reverse bias voltage of $-3$ V. The current represented by the dashed BBT curve neglects the strain contribution to the indirect gap $E^{ig}$ and to the effective mass in the transport direction. The solid BBT curve includes the full strain dependence. The Trap assisted currents are calculated with full strain dependence of $E_T$ and $\tau$ (grey), only $E_T$ (solid), only $\tau$ (dot-dashed) and strain independent $E_T$ and $\tau$ (dashed) (see Figs. \ref{tauvstr} and \ref{etvstr} for the strain dependence of $\tau$ and $E_T$). The blue points correspond to the experimentally measured current density.}
\end{figure}

Nowadays, strain of the order of 1\% is commonly used to increase the mobility of transistor channels. We explore the effects of such strains ($\sim 2$GPa in Si) on the leakage current, shown in Figs. \ref{jvstm1v} and \ref{jvstm3v} at voltages of -1V and -3V, respectively.
Strain affects BBT and TAT differently, as seen in Figure \ref{jvstm1v}. 
In this work, we calculate BBT entirely from first principles. We get excellent agreement between theory and experiment for the current density at zero strain. However, at stresses up to 180 MPa our theoretical model overestimates the response to strain compared to our experiment. This may be due to a variety of reasons, as explained earlier in the text. We find that the best quantitative agreement with experiment that retains the same linear behaviour of current vs stress is given by neglecting the stress dependence of the indirect gap $E^{ig}$ in eq. \ref{eqbbt} (see Fig. \ref{djAB135vst1-4vc-egi}). However, as explained earlier, we don't expect $E^{ig}$ to remain independent at stresses higher than 180 MPa. In our prediction of the BBT leakage current we consider both limiting factors: (i) complete stress dependence of all parameters and (ii) stress independence of $E^{ig}$. These two cases are shown in Figs. \ref{jvstm1v} and \ref{jvstm3v} as the red solid and dashed lines respectively. We observe that, except at small strains, the stress dependence of the current via the indirect gap is very strong. We must remark that the agreement of the model with experiments at low strains does not imply the stress dependence via $E^{ig}$ is negligible at higher strains. This stress dependence is very strong, and while it may be masked at lower strains by other factors, it should dominate at higher strains.

Notably, the full stress dependent BBT is most sensitive to compressive stress. 
Under enough compressive stress, BBT can become dominant over the other tunnelling mechanisms. 
Worryingly, the leakage current can increase by 4 to 80 times in TAT and BBT dominant devices under compressive stress at at $V=-1$V, and up to 20-150 times at higher voltages, where the current is dominated by BBT. Conversely, tensile stress affects the leakage to a lesser degree, and may reduce it at low stresses in purely BBT transport.

At this stage it is hard to predict the effect of strain on TAT. However, we can study the range of possible values under different regimes of extrapolation of the two unknown variables $E_T$ and $\tau$. These cases are shown in Figs. \ref{jvstm1v} and \ref{jvstm3v}. We explore the cases in which there is stress dependence in: (i) neither $E_T$ or $\tau$ (black dashed line), (ii) only $E_T$ (solid black), (iii) only $\tau$ (black dot-dashed) and (iv) both $E_T$ and $\tau$ (gray). Since both $E_T$ and $\tau$ are fitted by a cubic function of stress (see Figs. \ref{etvstr} and \ref{tauvstr}), this extrapolation to 2 GPa becomes very unrealistic. It may serve however as a limiting case. We expect the values of $E_T$ and $\tau$ to saturate at some point. Considering this, case (i) might give a reasonable expectation of the TAT current density at high stress. In general, we can expect TAT dominant leakage to increase less than 20 times under 2 GPa stress.

\section{CONCLUSION}
We combined experiment and theory to determine the effects of strain on the components of the tunnelling current in silicon diodes in reverse bias. 
We focused on diodes at three specific doping concentrations in which the current is dominated by either band-to-band tunnelling or trap-assisted tunnelling. 
The reverse bias current in the three diodes was measured at (100) uniaxial stress in the range of $\pm 180$ MPa. 
At the same time, we refined existing models of the band-to-band and trap-assisted tunnelling to account for the effects of strain. 
Whenever possible, we endeavoured to keep the models free from experimental parameters. 
This was possible entirely in the modelling of band-to-band tunnelling, where the electronic band structure and electron-phonon coupling parameters required for the calculation of the current have been calculated using first principles electronic structure methods.

The parameter-free description of trap-assisted tunnelling is possible to a degree. 
The details of the band structure and its response to strain can be treated in this way. 
However, the trap lifetimes and energies and their response to strain are not known.
We attempted to gain an understanding of this response, albeit crude, by fitting effective trap energies and lifetimes to our measurements vs. voltage and strain. 
The effective trap lifetimes and energies were the only fitting parameter in this otherwise parameter free model. 

The agreement between the model and experiment of the trap-assisted current is very good, in part due to the lifetimes and energies being fitted. The agreement is worse for the highly doped sample AB1, where other effects may be taking place at low voltages. 
The crossover from trap-assisted to band-to-band tunnelling with increasing voltage is well captured by the model, especially in AB3.

The agreement of the band-to-band component of the  current with strain is less satisfactory. 
Our model predicts a response of the current with strain that is 5 times larger than the one measured.
Having no parameters to fit, the model relies on the samples behaving as pure Si, and the strain being perfectly transferred across the sample. 
To understand the cause of the discrepancy, we modelled the band-to-band tunnelling current removing the strain dependency of each parameter in Eq. \ref{eqbbt} separately. 
We found the only effect consistent with measurement to be the lack of a strain dependence of the band gap. 
This may be due to the large doping concentration and small strains considered, and warrants further study using higher strains and a model explicitly considering dopant states.

We used our theoretical model to venture a prediction of the tunnelling current at higher uniaxial stresses of up to 2 GPa, as typically found in stress enhanced transistor channels. 
These stresses are beyond the range of validity of our fit of the effective trap lifetimes and energies. 
However, we considered four likely limiting scenarios of the behaviour of these trap variables based on the small strain fit: extrapolation of the strain dependence of both, either or neither the trap energies and lifetimes. 

Also, while our band-to-band model is parameter-free and is in excellent agreement with our measurements at zero strain, the low strain measurements that hint at a strain independent band gap may be indicating there are effects not considered in the model. 
If these effects are due to the high doping charge, it may saturate at a certain stress, beyond which there is a strain dependence of the gap. 
We therefore considered the limiting cases of both a strain-independent and strain-dependent energy gap in the band-to-band tunnelling current. 

We predict that strain can have a very large effect on the leakage current, especially in band-to-band dominated samples. 
At low voltages compressive strain can switch a trap-assisted-tunnelling to a band-to-band dominated current.
If, however, the strain dependence of the band gap is small or absent as suggested by the measurements at small strain, band-to-band tunnelling may be largely insensitive to strain. 

Trap-assisted-tunnelling is less sensitive to strain than band-to-band tunnelling, provided we include the strain dependence of the band gap. 
The trap-assisted leakage current increase due to strain is very sensitive to the limiting case for the trap energies and lifetimes considered. 
The most extreme case is that considering the trap lifetimes and energies independent of strain in the trap dominated sample AB5. 
In this case, the current increases by about 20 times with a compressive stress of 2 GPa. 
The other liming cases produce a lower increase, or in the case of only strain dependent energies, a severe decrease of the leakage current. 
We expect the case of strain independent effective trap energies and lifetimes, and strain dependent lifetimes to be the closest to reality, for the following reasons. 
Generally, there are multiple trap states that have different energies and lifetimes. 
Our model simplifies the effect of all these traps into that of one, effective trap state. 
At high strains the shift in the average energy of multiple trap states is likely to be less than an effective state, as states that move out of the tunnelling window of energies are replaced by those entering it from the opposite extreme. 
The lifetimes are more likely to change if the stress environment is similar for all traps present.

Band-to-band and trap-assisted tunnelling differ in their response to strain in another important way: doping strongly affects the functional response of trap-assisted-tunnelling, while it only changes the scale of band-to-band tunnelling. 
Electric field strength only affects the scale of both tunnelling mechanisms.

In this work we provided a first step to understand the effects of strain on the leakage current of trap and band-to-band tunnelling dominated \it{pn} junctions. 
Our parameter-free band-to-band model reproduces the current-voltage characteristics of our measurements, but overestimates the response to strain. 
We believe that the high doping is affecting the band gap response to strain. 
To settle this question, measurements at higher strains, different strain configurations and a more comprehensive model are required.

We also determined that trap energies and lifetimes are affected by strain. 
However, the nature of this dependence is not yet clear. This question should be addressed by an atomistic study of the response to strain of known trap states.

Finally, the tunnelling processes studied in this work are the same driving the current in TFET devices. The results of this work regarding the interplay of strain, voltage, doping and the transition from trap- to band-to-band assisted tunnelling should be useful in the design of TFETs with improved sub-threshold slopes.


\section{Acknowledgements}

FMA would like to acknowledge the SFI award 12/IA/1601 for funding this work.

\bibliography{sileakage}
\end{document}